# Energy Efficient and Delay Aware Vehicular Edge Cloud

**Amal A. Alahmadi, T. E. H. El-Gorashi, and Jaafar M. H. Elmirghani**
*School of Electronic & Electrical Engineering, University of Leeds, LS2 9JT, United Kingdom*
*e-mail: {elaaal t.e.h.Elgorashi, j.m.h.elmirghani}@leeds.ac.uk*

**ABSTRACT**
Vehicular Edge Clouds (VECs) is a new distributed processing paradigm that exploits the revolution in the processing capabilities of vehicles to offer energy efficient services and improved QoS. In this paper we tackle the problem of processing allocation in a cloud-fog-VEC architecture by developing a joint optimization Mixed Integer Linear Programming (MILP) model to minimize power consumption, propagation delay, and queuing delay. The results show that while VEC processing can reduce the power consumption and propagation delay, VEC processing can increase the queuing delay because of the low data rate connectivity between the vehicles and the data source nodes.
**Keywords**: vehicular edge computing; processing allocation; power consumption; propagation delay; queuing delay.

## 1. INTRODUCTION

The recent increase in the number of connected devices is accompanied by a rapid growth in networks traffic and data centers load [1]. This growth has led to a significant increase in the power consumption of the communication networks and data centers. Research efforts have been focused on developing energy efficient architectures for cloud data centres and core networks [2]–[5]. Different techniques and technologies are considered to improve network energy efficiency including virtualization [6]–[8], network architecture design and optimization [9]–[13], content distribution [14]–[16], big data [17] - [20], network coding [21], [22] and renewable energy [23]. Decentralized architectures have also been proposed to integrate distributed edge servers to mitigate the traffic burden on central data centers [23] - [25].

The revolution in the processing capabilities of vehicles promises a new of paradigm of distributed processing referred to as Vehicular Edge Cloud (VEC). Recent research efforts have demonstrated the potential of vehicles clustered in a parking lot [26] and vehicles clustered in roads [27] to form VECs. By offloading some of the cloud burden to the VEC. The network power consumption can be reduced and QoS can be maintained for delay sensitive applications. Our previous work in [28] optimized the processing allocation in a cloud-fog-VEC architecture to minimize the total power consumption considering different allocation strategies and task requirements. The effort in [29] developed a flexible offloading algorithm to optimize the processing placement in central clouds, cloudlet servers and vehicles to reduce the computational overhead of the smartphone given its limited processor capabilities. The work also assessed the power consumption and response time of the processed tasks. The authors in [30] presented another attempt to minimize energy consumption by optimizing the task-offloading decisions in a VEC framework. Less attention was dedicated to joint optimization problems with more than one objective. The authors in [31] developed a model to minimize network delay, minimize power consumption, and maximize the availability of the processing resources. Joint optimization models are proposed to minimize service delay and quality loss in [32] and minimize response delay and maximize processing success in [33].

In this paper, we extend our previous work in [28], where the focus was on minimizing the total power consumption, to consider joint minimization of the total power consumption, propagation delay, and queuing delay in a cloud-fog-VEC architecture. The remainder of this paper is organized as follows: In Section 2, we present the cloud-fog-VEC architecture and discuss the joint optimization model. The optimization results are presented and discussed in Section 3. Finally, Section 4 concludes the paper.

## 2. JOINT OPTIMIZATION OF POWER CONSUMPTION AND DELAY FOR THE CLOUD-FOG-VEC ARCHITECTURE

Figure 1 shows the proposed cloud-fog-VEC architecture. The architecture connectivity is provided by a GPON at the access network, a metro network and an IP over WDM network at the core network. Processing is provided by processing nodes (PNs) at five layers: central cloud (CC), metro fog (MF), OLT fog (LF), ONU fog (NF) and vehicular edge cloud (VEC). The VEC layer consists of vehicular nodes (VNs) clustered in car parks, charging stations or road intersections. Processing tasks requested by nearby devices are sent to an access point (AP) which allocates the tasks generated to the PNs based on the available capacity of PNs and other criteria. The AP is connected to the VNs using WiFi and to other PNs using fiber links. We formulate the processing resources allocation problem as a MILP model to minimize the total power consumption, propagation delay and queuing delay from processing tasks.

The power consumption is composed of the processing power consumption and networking power consumption. The processing power consumption accounts for both vehicle on-board units and servers at fog and cloud nodes, and the networking devices within the fog and cloud nodes if any. The networking power consumption is composed of the GPON, metro network and core network power consumption.

The distances illustrated in Figure 1 (shown in km) are used to calculate the propagation delay as given below

$$Propagation\ Delay = \frac{D}{\Delta RI\ \mathbb{C}} \qquad (1)$$

where $D$ is the distance between any two networking devices in Figure 1, $\mathbb{C}$ is the speed of light (299,792 km/s), and $\Delta RI$ is a factor that accounts for the slower propagation of light in the fibre optic link and is the receiprocal of the refractive index of glass ($\Delta RI = \frac{2}{3}$).

The queuing delay is modelled for each networking node as an M/M/1 queue with one server, where arrivals follow a Poisson process and the service rate is negative exponentially distributed. The aggregated traffic delivered to each node and the network device capacity define the arrival rate ($\lambda$) and service rate ($\mu$), respectively, and are used to determine the queueing delay as given below;

$$Queuing\ Delay = \frac{1}{\mu - \lambda} \qquad (2)$$

We have considered in this work delay at the packet level. We used the Ethernet maximum packet size of 1500 bytes and therefore expressed the arrival data rates as packets per second and expressed the service rates (transmission rates) in packets per second.

The AP works at two different service rates; 10 Gb/s for the interface to the GPON and 1 Gb/s for the wireless interface to the VEC layer. The core nodes and the data centres attached to them work at 40 Gb/s. Other network devices are assumed to run at 10 Gb/s service rate (GPON).

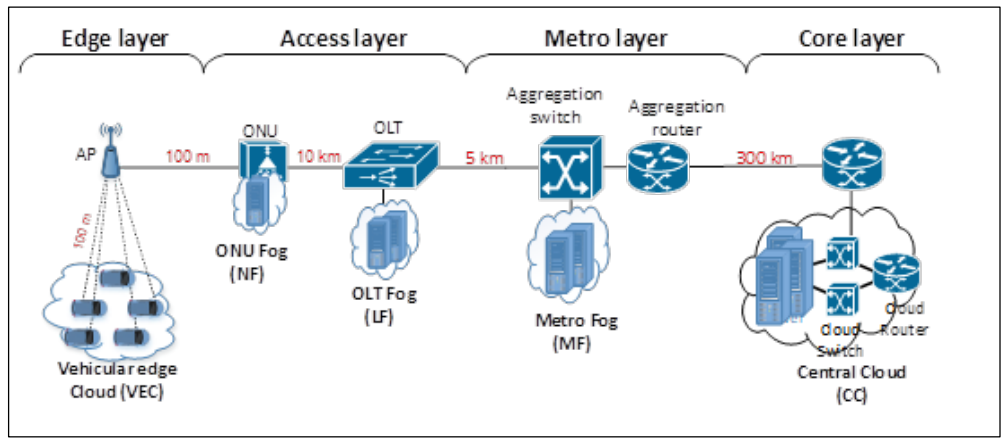

Figure 1. Integrated Vehicular Cloud Architecture

The tasks allocation is optimized using a mixed integer linear programming (MILP) model with the objective of minimizing the total power consumption, propagation delay, and queuing delay.

The joint objective function is defined as:

$$\text{Minimize} \qquad \alpha P + \beta R + \gamma Q \qquad (3)$$

where $P$, $R$, and $Q$ are the total power consumption, the propagation delay, and the queuing delay, respectively. The weighting factors $\alpha$, $\beta$, and $\gamma$ are used to scale the terms so that they are comparable in magnitude, and to accommodate the units in the objective function. Therefore, $\alpha$ is a unitless factor, and $\beta$ & $\gamma$ have units of $\frac{Watt}{sec}$.

Three scenarios are considered as follows:

1) Minimizing the power consumption and traffic propagation delay jointly, by setting the values of $\alpha$ to 1, $\gamma$ to zero, and $\beta$ to a value that ensures equal importance of the power consumption and propagation delay.

2) Minimizing the power consumption and traffic queuing delay jointly, by setting the values of $\alpha$ to 1 and $\beta$ to zero, and $\gamma$ to a value that ensures equal importance of the power consumption and queuing delay.

3) Minimizing the power consumption, traffic propagation and queuing delay jointly, by setting the values of $\alpha$ to 1, and the values of $\beta$ and $\gamma$ to values that ensure $\alpha P = \beta R = \gamma Q$.

## 3. RESULTS AND DISCUSSION

We evaluate the processing allocation under the three scenarios described above in a cloud-fog-VEC architecture with 8 vehicles of homogeneous processing capabilities provided by their on board units (OBUs), and one server at each of the other PNs. We assessed ten tasks that represent a range of processing requirements (100–1000 MIPS for each task) and a range of traffic demands (10–100 Mb/s for each task) assuming a task can only be processed by a single PN, ie task splitting is not permitted.

Table 1 summarizes the processing capacity and power consumption of components in the different layers of the architecture.

*Table 1. Network components capacity and power consumption parameters*

| Network Components | Maximum capacity | Maximum power (W) | Idle power (W) | Network Components | Maximum capacity | Maximum power (W) | Idle power (W) |
|---|---|---|---|---|---|---|---|
| Central cloud (CC) server [34] | 144k MIPS | 115 | 69 | Vehicular node (VN) processor [35] | 3.2k MIPS | 10 | 6 |
| CC router port [36] | 40Gb/s | 25 | 22.5 | VN Wi-Fi adapter [37] | 72.2 Mb/s | 2.5 | 1.5 |
| CC switch [38] | 600 Gb/s | 460 | 414 | Core router port [39] | 40Gb/s | 638 | 574.2 |
| Metro Fog (MF) server [40] | 88k MIPS | 85 | 51 | Metro router port [36] | 40Gb/s | 25 | 22.5 |
| OLT fog (LF) server [41] | 54.4k MIPS | 85 | 51 | Metro switch [42] | 1800Gb/s | 500 | 450 |
| MF&LF router port [43] | 40 Gb/s | 13 | 11.7 | OLT [44] | 1920 Gb/s | 50 | 45 |
| MF&LF switch [45] | 200 Gb/s | 245 | 220.5 | ONU [46] | 10Gb/s | 15 | 13.5 |
| ONU fog (NF) processor [47] | 6k MIPS | 15 | 9 | AP [48] | 1 Gb/s | 11 | 4.8 |

Figures 2 shows the total power consumption versus the total traffic generated by the tasks for the three scenarios under consideration. The results are relatively comparable for the three scenarios. The only difference is noticed with the minimized power and propagation delay scenario with low traffic (100-300 Mb/s), where the results indicate lower power consumptions compared to the other two scenarios. This is due to the fact that all the tasks at 100-300 Mb/s are served by the VEC, as seen in Figure 5, which is considered the most energy efficient PN. The jump in the curves in the three minimization functions is attributed to moving the allocation at higher traffic to less efficient PNs that can support the traffic. For instance, at 800 Mb/s (in Figure 2), the jump is a result of moving the allocation from VN and NF to NF and LF, as seen in Figure 5. This is because the connectivity between the AP and a VN cannot support the 80 Mb/s connection required by a task, and the NF cannot accommodate all the generated tasks.

Figure 3 shows the average propagation delay versus the total traffic generated by the tasks for the three scenarios. It is worth mentioning that the queuing delay value is expressed in sec/packet, where we used the Ethernet MTU packet size of 1500 Byte. As it reflects the distance between the AP and the activated PNs, the propagation delay will be reduced for tasks allocated to the nearest PNs (based on the estimated distances shown in Figure 1). Since the power consumption results (in Figure 2) indicated that VEC is the most energy efficient PN, comparable results for the propagation delay are evident in Figure 3. The minimum average propagation delay is achieved when minimizing both power and propagation delay at low traffic, as tasks are allocated to VEC. The jump in Figure 2 at 800 Mb/s is noticed again in Figure 3. This can be explained by the limited VNs connection, described previously, leading to the activation of LF which is 10 km away from the ONU.

Figure 4 shows the average queuing delay versus the total traffic generated by the tasks for the same three scenarios. As described earlier, the queuing delay of a node is influenced by the service rate of this node. Hence, allocating tasks to central cloud and fixed fog nodes which requires traversing network nodes with a high service rate results in lower queuing delay, compared to allocating tasks to the VNs connected to the low service rate AP wireless interface.

Considering queuing delay in the optimization resulted in avoiding VNs as shown in Figure 5. For the power and propagation delay minimized scenario, low demand tasks were allocated to VNs (as seen in Figure 5) which results in a high average queuing delay (13–17 μs). As the traffic increases, tasks were allocated to NF (as VNs became insufficient) and the average queuing delay dropped by 85% (from 17 μs to 2.5 μs). It saturates at this value as long as the NF remains the most power efficient placement. However, the average queuing delay shows a slight increase at 700 Mb/s as the tasks are allocated to two PNs (i.e. NF and VN) due to NF insufficient capacity.

With the increase in the generated traffic, beyond 700 Mb/s, the tasks are allocated to NF and LF (passing through ONU and OLT). This results in reducing the average queuing delay, compared to the case at 700 Mb/s.

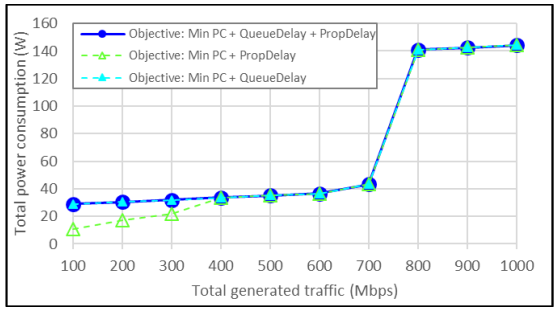

Figure 2. Total power consumption

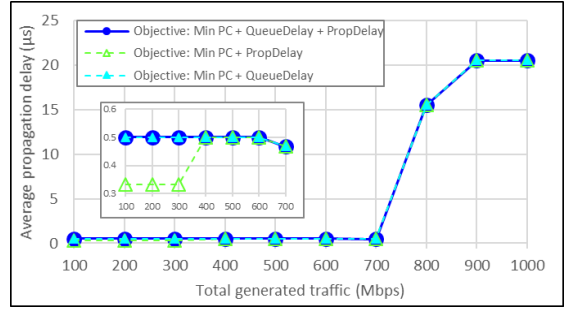

Figure 3. Average propagation delay

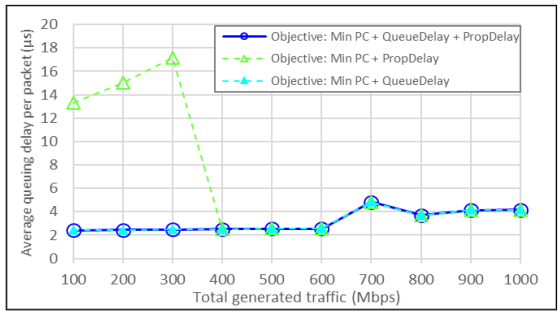

Figure 4. Average queuing delay

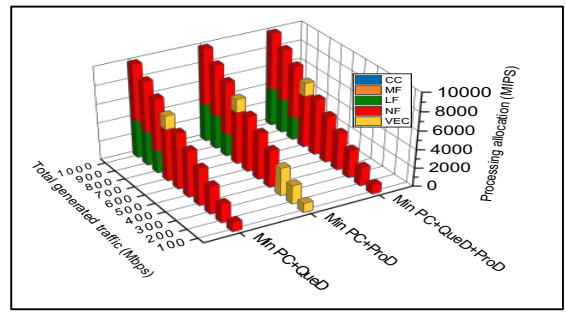

Figure 5. Processing allocation in each PN

## 4. CONCLUSIONS

In this paper, we have investigated the joint minimization of power consumption, propagation delay and queueing delay when allocating processing tasks to PNs in a cloud-fog-VEC architecture. The processing resources allocation problem was formulated as a MILP model to minimize the total power consumption, propagation delay and queuing delay. Our results show that VEC processing yields the minimum power consumption and propagation delay. However, the queuing delay becomes the limiting factor of VEC processing as the AP wireless interface connecting the source nodes to the VNs operates at a low data rate compared to the GPON connecting the source nodes to the fog and cloud PNs. Future work can examine APs with different data rates to reduce the queuing delay of VEC. Additional delay components can be considered in the optimization including processing delay and transmission delay alongside propagation and queuing delay.

## ACKNOWLEDGEMENTS

The authors would like to acknowledge funding from the Engineering and Physical Sciences Research Council (EPSRC), through INTERNET (EP/H040536/1), STAR (EP/K016873/1) and TOWS (EP/S016570/1) projects. All data are provided in full in the results section of this paper.